\documentclass[prl,showpacs,superscriptaddress,twocolumn,preprintnumbers,amsmath,amssymb]{revtex4-1}

\usepackage[dvipdfmx]{graphicx}
\usepackage{dcolumn}
\usepackage{bm}
\usepackage{amsmath}
 
\graphicspath{{graphics/}}  
\usepackage{color}
\usepackage{cancel}
\usepackage{ulem}
\definecolor{mymauve}{rgb}{0.58,0,0.82}
\definecolor{mygreen}{rgb}{0,0.5,0.25}

\begin{document}


\title{Strongly coupled single quantum dot-cavity system\\integrated on a CMOS-processed silicon photonic chip}

\author{A.~Osada}
\email{alto@iis.u-tokyo.ac.jp}
\author{Y.~Ota}
\affiliation{Institute for Nano Quantum Information Electronics, The University of Tokyo, Meguro-ku, Tokyo 153-8505, Japan.}
\author{R.~Katsumi}
\affiliation{Institute of Industrial Science (IIS), The University of Tokyo, Meguro-ku, Tokyo 153-8505, Japan}
\author{M.~Kakuda}
\affiliation{Institute for Nano Quantum Information Electronics, The University of Tokyo, Meguro-ku, Tokyo 153-8505, Japan.}
\author{S.~Iwamoto}
\affiliation{Institute for Nano Quantum Information Electronics, The University of Tokyo, Meguro-ku, Tokyo 153-8505, Japan.}
\affiliation{Institute of Industrial Science (IIS), The University of Tokyo, Meguro-ku, Tokyo 153-8505, Japan}
\author{Y.~Arakawa}
\affiliation{Institute for Nano Quantum Information Electronics, The University of Tokyo, Meguro-ku, Tokyo 153-8505, Japan.}

\date{\today}

\begin{abstract}
Quantum photonic integrated circuit (QPIC) is a promising tool for constructing integrated devices for quantum technology applications. In the optical regime, silicon photonics empowered by complementary-metal-oxide-semiconductor (CMOS) technology provides optical components useful for realizing large-scale QPICs. Optical nonlinearity at the single-photon level is required for QPIC to facilitate photon-photon interaction. However, to date, realization of optical elements with deterministic( i.e., not probabilistic) single-photon nonlinearity by using silicon-based components is challenging, despite the enhancement of the functionality of QPICs based on silicon photonics. 
In this study, we realize for the first time a strongly coupled InAs/GaAs quantum dot-cavity quantum electrodynamics (QED) system on a CMOS-processed silicon photonic chip. The heterogeneous integration of the GaAs cavity on the silicon chip is performed by transfer printing. 
The cavity QED system on the CMOS photonic chip realized in this work is a promising candidate for on-chip single-photon nonlinear element, which constitutes the fundamental component for future applications based on QPIC, such as, coherent manipulation and nondestructive measurement of qubit states via a cavity, and efficient single-photon filter and router.
\begin{description}
\item[PACS numbers]
\end{description}
\end{abstract}

\maketitle



Quantum photonic integrated circuits (QPICs)~\cite{OBrien2009, Silverstone2016}, comprised of various kinds of quantum optical elements assembled on a single chip, are intensively studied as promising tools for the implementation of quantum devices for quantum technology applications ~\cite{NielsenChuang,Georgescu2014,Gisin2002,Degen2017}. Various elements commonly used for quantum and classical optics devices, such as, beam splitters and modulators, are well-developed based on a silicon photonics platform~\cite{Soref2006,Dai2012}, which employs the complementary-metal-oxide-semiconductor (CMOS) process. However, in silicon photonics, elements with single-photon nonlinearity have not yet been implemented despite their ability to realize photon-photon interaction required for applications, such as, universal quantum computation. As a candidate for the implementation of such elements, a system involving an optical cavity with a single two-level emitter, which is called the cavity quantum electrodynamics (QED) system, is of significant interest. The cavity QED system has been investigated with various two-level emitters~\cite{Raimond2001, Aoki2006, Steiner2013,Park2006,Xiang2013,Yoshie2004,Reithmaier2004}.  Strong modification of the electromagnetic environment surrounding the emitter by using cavities and waveguides has allowed researchers to achieve various quantum optical elements, e.g., efficient and fast single-photon sources~\cite{Michler2000,Kuhn2002,Englund2010,Houck2007}, single-photon switches and routers~\cite{Shomroni2014, Volz2012, Hoi2011}, and Fock-state filters~\cite{Tiecke2014,Javadi2015}. 

Among the cavity QED systems, the one with a semiconductor quantum dot (QD) is promising to be integrated into QPIC, because of its well-developed electrical controllability and availability of telecommunication-compatible emission wavelengths that might be suitable for silicon photonics.  Some heterogeneously integrated QD light sources without optical cavities~\cite{Zadeh2016, Elshaari2017, Kim2017} have been implemented by using methods, such as, wafer bonding on silicon-based photonic circuits.  However, cavity QED-based elements combined with CMOS photonic circuitry are yet to be realized despite their considerable potential to provide highly functional single-photon devices to be integrated with QPIC. Even without the use of the CMOS technology, only a limited number of realizations of cavity QED systems on Si platform~\cite{Davanco2017,Luxmoore2013} have been reported so far, where Ref.~\cite{Davanco2017} utilized a silicon nitride system and Ref.~\cite{Luxmoore2013} implemented cavity QED systems on a hybrid substrate without any silicon-based photonic elements. The main obstacle here is the complicated process optimization required to simultaneously fabricate a nanoscale compound-semiconductor cavity and low-loss silicon photonic element. This issue can be resolved by transfer printing~\cite{Menard2004, Meitl2006, Yoon2015} that allows pick-and-place heterogeneous integration of individually fabricated elements, which is suitable for integrating nanoscale cavity QED systems into CMOS-processed silicon photonic circuits. With such a method, a cavity QED system in the strong coupling regime on a CMOS photonic chip is feasible, which is particularly significant not only in view of efficient, on-chip coherent manipulation and non-destructive readout of QD states by using the cavity~\cite{Brune1990, Wallraff2004}, but also because of its scalability and functionality aided by the CMOS technology.
   

\begin{figure}[t]
\includegraphics[width=8.6cm]{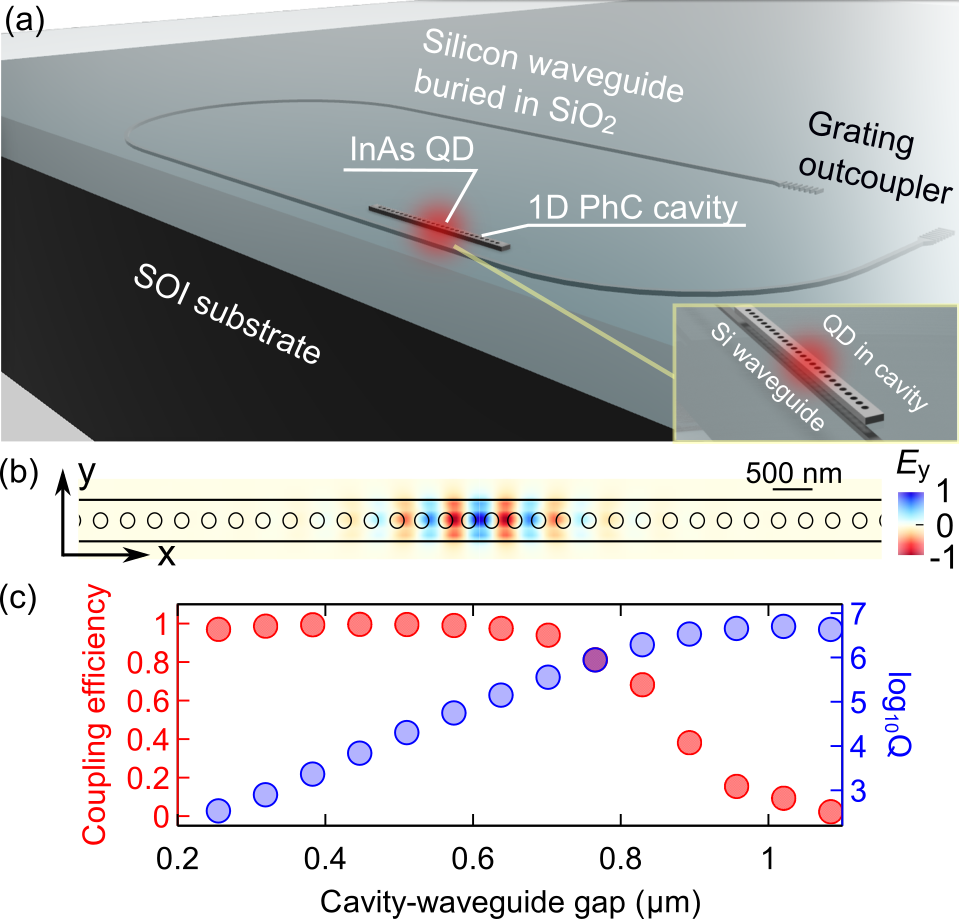}
\caption{\label{Fig1} (a) Schematic illustration of the device implementing strongly coupled quantum dot-cavity system integrated on a CMOS-processed silicon photonic circuit. (b) Calculated distribution of the $y$-component of the electric field exhibited by the fundamental mode of the one-dimensional photonic crystal cavity. The field is normalized to the maximum value. (c) Numerically evaluated quality factor (blue, right axis) of the fundamental mode and the fraction of light coupled into the silicon waveguide (red, left axis) as functions of the distance between the cavity and waveguide.}
\end{figure}

In this Letter, we demonstrate a strongly coupled single QD-photonic crystal (PhC) cavity system integrated on a CMOS-processed silicon photonic circuit by transfer-printing method.  A one-dimensional PhC cavity containing InAs/GaAs QDs is transfer-printed on top of a silicon waveguide on a silicon-on-insulator (SOI) chip that is fabricated in a CMOS-process foundry. The device is designed to offer near-unity coupling efficiency to the waveguide and high quality factor of the cavity.  Micro-photoluminescence spectroscopy reveals the vacuum Rabi splitting between the QD and the cavity, which indicates that the QD-cavity coupled system is in the strong coupling regime.  We further analyze the spectra to obtain the coupling strength as $g_0 = 69$~$\mu$eV and observe the interchanged peak heights and linewidths between the upper and lower polaritons, which substantiate the strong coupling.  From a grating outcoupler terminating the waveguide, we observe the vacuum Rabi splitting that confirms the waveguide-coupling of the system.  

\begin{figure}[t]
\includegraphics[width=8.6cm]{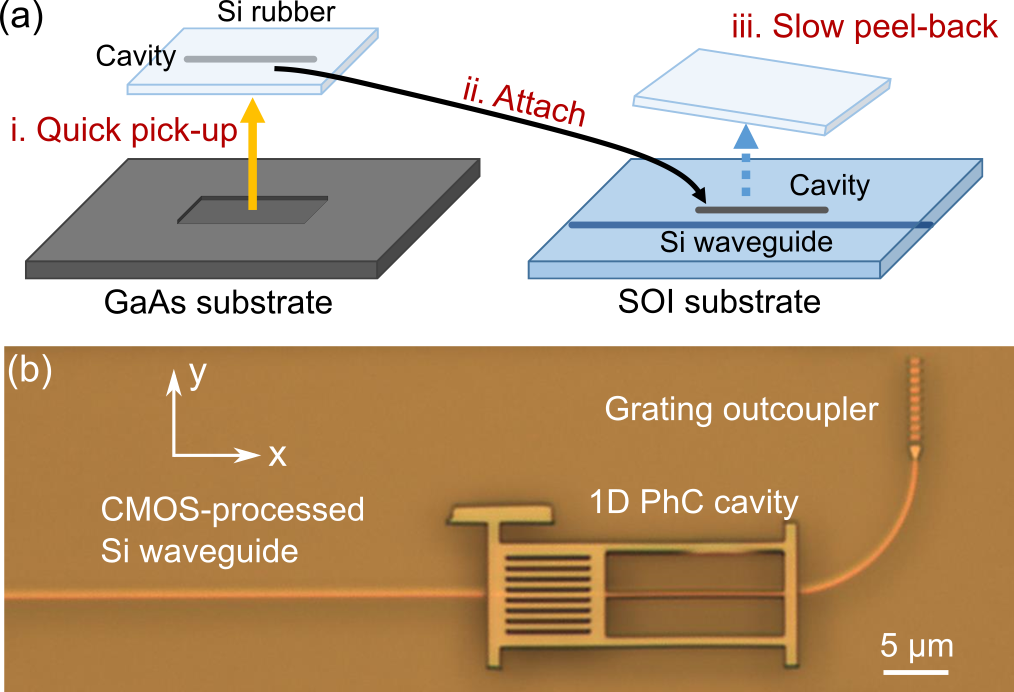}
\caption{\label{Fig2} (a) Schematic of the transfer-printing process. An air-bridged one-dimensional photonic crystal cavity is picked up by a silicon rubber from a GaAs chip and then attached onto a silicon-on-insulator chip on top of a waveguide.  By slowly peeling back the silicon rubber, the cavity is left on the waveguide.  The whole procedure is monitored by an optical microscope.  (b) Optical micrograph of the fabricated device where a one-dimensional photonic crystal cavity is transfer-printed on a CMOS-processed silicon waveguide, which is terminated by grating outcouplers.}
\end{figure}

\begin{figure*}[t]
\includegraphics[width=16cm]{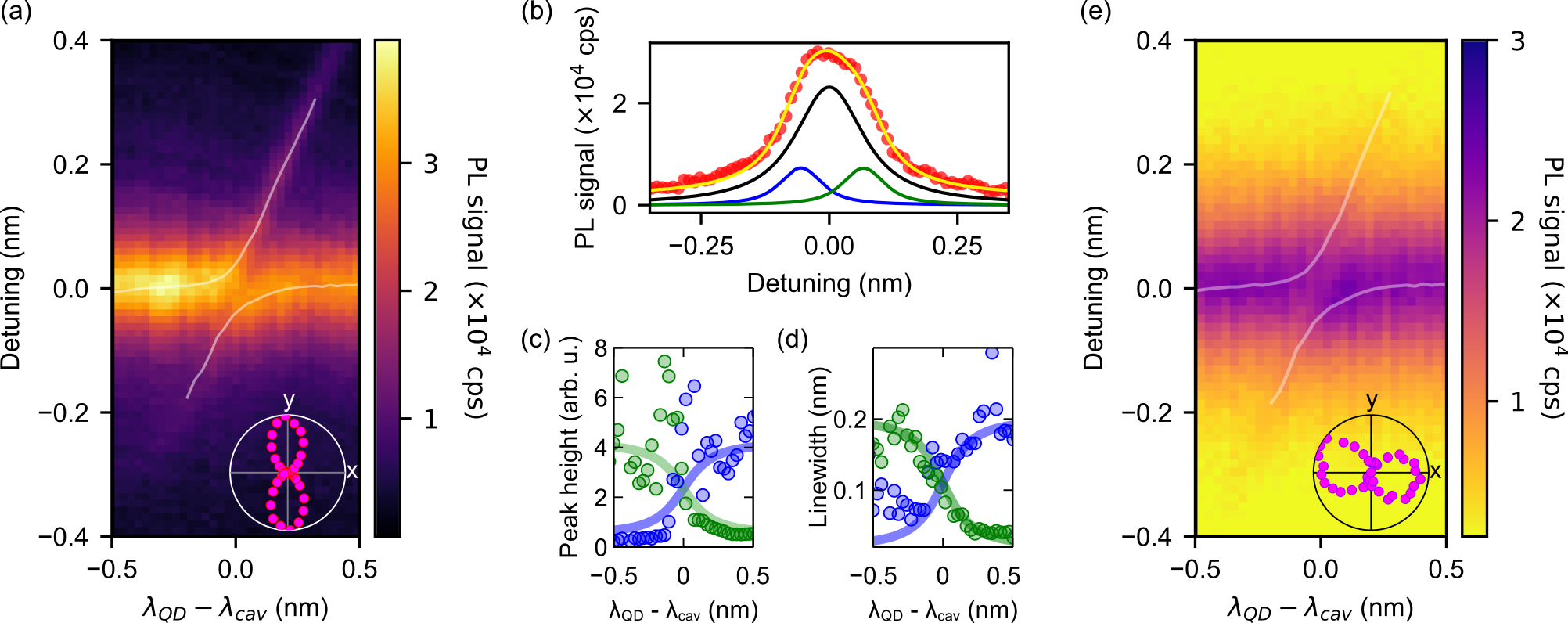}
\caption{\label{Fig3} (a) Color plot of PL spectra with the temperature of the device ranging from 10 K to 35 K. The wavelength of the quantum dot (QD) $\lambda_{\mathrm{QD}}$ relative to that of the cavity $\lambda_{\mathrm{cav}}$, as shown in the horizontal axis. The vertical axis represents the wavelength detuning from $\lambda_{\mathrm{cav}}$. The white curves indicate the positions of peaks originating from the upper and lower polaritons extracted by fitting. The polarization of the detected signal when $\lambda_{\mathrm{QD}}-\lambda_{\mathrm{cav}}=0.5$~nm is shown in the bottom right. (b) Spectrum at $\lambda_{\mathrm{QD}} = \lambda_{\mathrm{cav}}$~nm (red) analyzed by multi-peak fitting (yellow) with the components of the upper (blue) and lower (green) polaritons and the bare cavity (black) plotted together. (c), (d) Extracted (c) peak heights and (d) linewidths of upper (blue) and lower (green) polaritons with the solid curves indicating the simulated behavior. (e) Color plot of PL signals obtained from the grating outcoupler, plotted in the same way as Fig.~\ref{Fig3}(a). The white curves indicate the positions of the upper and lower polaritons extracted in Fig.~\ref{Fig3}(a) as a guide to the eye. The polarization of the detected signal when $\lambda_{\mathrm{QD}}-\lambda_{\mathrm{cav}}=0.5$~nm is shown in the bottom right.}
\end{figure*}

The device fabricated and investigated in this study is a one-dimensional (1D) PhC cavity~\cite{Deotare2009, Kuramochi2010, Gong2010, Ohta2011} coupled to a silicon wire waveguide buried in silicon dioxide, where the cavity strongly couples to a single QD embedded in it.  A schematic illustration of the device is shown in Fig.~\ref{Fig1}(a).   The PhC cavity is formed by a nanobeam with almost equidistantly aligned air holes, where the intervals between the adjacent air holes are modulated in the central region containing 10 air holes~\cite{Ohta2011}. The ratio of the radius $r$ of the holes to the lattice constant $a$ is $r/a = 0.26$. The nanobeam has a thickness of 180~nm and width of 471~nm.  The distribution of the electric field in the $y$-direction of the fundamental cavity mode is calculated by finite difference time domain (FDTD) method, and is shown in Fig.~\ref{Fig1}(b). The unloaded quality factor is $6\times10^6$ and the mode volume is $0.56\,(\lambda/n)^3$, where $\lambda = 1.16$~$\mu$m is the resonant wavelength and $n = 3.4$ the refractive index of GaAs. Here, the unloaded (loaded) quality factor refers to that of the cavity on a silicon dioxide layer in the absence (presence) of the waveguide coupling. Underlying is a silicon wire waveguide buried in a silicon dioxide layer. Its thickness and width are, respectively, 210 and 250 nm so that the waveguide supports only a single transverse-electric (TE) mode at the wavelength of around $1160$~nm.  This also approximately fulfills the phase-matching condition between the waveguide and cavity~\cite{Katsumi2018} for efficient coupling between them. The waveguide is terminated by grating outcouplers for collecting the signal introduced to the waveguide from the cavity.

The cavity and waveguide are separated by a silicon dioxide layer.  Coupling efficiency, i.e., the ratio of light leaking from cavity mode into the silicon waveguide to total leakage, is controlled by the gap between the cavity and waveguide in this design.  As the cavity-waveguide gap becomes smaller, the loaded quality factor of the cavity mode decreases while the coupling efficiency approaches unity [see FDTD calculation results in Fig.~\ref{Fig1}(c)].  The designed value of the gap, $570$~nm, results in a coupling efficiency of 99~\% and loaded quality factor of $4.8\times10^4$, which are advantageous together with the small mode volume of the cavity for the experiment conducted in this study.  The high coupling efficiency in this design is attributed to the large unloaded quality factor of the cavity, as obtained in the numerical calculation.  

The device is fabricated as follows.  First, air-bridged 1D PhCs supported by square frames are prepared by electron-beam lithography, and dry and wet etching. The square frame surrounding the PhC cavity protects the cavity from bending and breaking and facilitates the picking-up process in transfer printing. The silicon wire waveguides on an SOI substrate are fabricated in a CMOS-process foundry and initially capped with a $1.8$-$\mu$m-thick silicon dioxide layer formed by chemical vapor deposition.  The silicon dioxide layer on the top of the silicon waveguides are thinned down to $573$~nm by dry etching.  With these constituent elements, the device is fabricated by transfer printing, which is achieved by picking-up a 1D PhC cavity using a silicon rubber, attaching the cavity on top of the waveguide, and peeling-back the silicon rubber to leave the cavity on the SOI substrate.  It has been shown that the PhC cavities can be transfer-printed with an alignment precision within $100$~nm and the quality factor maintained as high as $\sim 10^4$~\cite{Katsumi2018,AO2018}.  These procedures are monitored with an optical microscope and are schematically illustrated in Fig.~\ref{Fig2}(a). An optical micrograph of the fabricated device is shown in Fig.~\ref{Fig2}(b), from which it is deduced that the precision of alignment of the cavity with respect to the waveguide is below $100$~nm.

The fabricated device is evaluated by low-temperature micro-photoluminescence (PL) spectroscopy with $785$~nm-wavelength; here, continuous-wave laser as an excitation laser impinges on the cavity with an optical power of 9.8~$\mu$W through a $\times$50 objective lens.  First, we evaluate the device by collecting PL signals radiated above the cavity.  Figure~\ref{Fig3}(a) displays the series of PL spectra for variable wavelength detuning between the quantum dot ($\lambda_{\mathrm{QD}}$) and the fundamental mode of the cavity ($\lambda_{\mathrm{cav}}$). The wavelengths $\lambda_{\mathrm{QD}}$ and $\lambda_{\mathrm{cav}}$ are simultaneously tuned by varying the temperature of the device from 10 K to 35 K, where $\lambda_{\mathrm{QD}}$ varies more than $\lambda_{\mathrm{cav}}$.  In Fig.~\ref{Fig3}(a), the vertical axis represents the wavelength of the QD relative to the resonant wavelength of the cavity, $\lambda_{\mathrm{cav}} \sim 1\,155$~nm. The cavity mode exhibits a loaded quality factor of $8\,000$ that corresponds to the total decay rate of the cavity, $\kappa_{\mathrm{tot}} \sim 2\pi \times 32$~GHz, which is evaluated when $\lambda_{\mathrm{QD}} - \lambda_{\mathrm{cav}} = 0.5$~nm.  

When $\lambda_{\mathrm{QD}}=\lambda_{\mathrm{cav}}$, an avoided crossing between the QD and cavity is observed.  The white curves in Fig.~\ref{Fig3}(a) indicate the positions of the peaks of upper- and lower-polaritons extracted by fitting, which also confirms the avoided crossing behavior.  The bottom-right inset displays the polarization of the observed signal above the cavity when $\lambda_{\mathrm{QD}}-\lambda_{\mathrm{cav}}=0.5$~nm, see Fig.~\ref{Fig2}(b) for the definition of the directions $x$ and $y$. The $y$-polarized signature agrees with the characteristics of the polarization of the cavity; hence, it is verified that the observed signal originates from the emission above the cavity. The spectrum at $\lambda_{\mathrm{QD}} = \lambda_{\mathrm{cav}}$ is shown in Fig.~\ref{Fig3}(b) together with the result of fitting (yellow solid line) with three Voigt functions for the peaks of upper polariton (blue), lower polariton (green) and residual signal of the cavity (black)~\cite{Hennessy2007}.  The vacuum Rabi splitting is deduced to be $122$~$\mu$eV, from which the coupling strength between the QD and cavity is determined as $g_0 = 69$~$\mu$eV.  For the coupling strength $g_0 = 2\pi \times 17$~GHz, decay rate of the cavity $\kappa_{\mathrm{tot}} = 2\pi \times 32$~GHz and that of the QD $\gamma_{\mathrm{QD}} \ll \kappa_{\mathrm{tot}}$, the system exhibits $g_0 > \kappa_{\mathrm{tot}}/4, \gamma_{\mathrm{QD}}/4$ that fulfills the condition of strong coupling.  In Fig.~\ref{Fig3}(c) and (d), respectively, peak heights and linewidths of the upper (blue) and lower (green) polaritons are plotted against $\lambda_{\mathrm{QD}}-\lambda_{\mathrm{cav}}$ with the solid curves representing the results of analytical simulation based on the Jaynes-Cummings model~\cite{JC1963} adopting relevant experimental parameters for guide to the eye.  Both plots indicate the interchanging spectral properties between the upper/lower polaritons over the avoided crossing, which further establishes the strong coupling between the QD and cavity in this device.


Next, the waveguide coupling of the cavity is examined by shining the excitation laser on the cavity and collecting PL signal from the grating outcoupler. Here, the irradiated optical power is 14~$\mu$W.  The temperature of the device is again tuned from 10 K to 35 K to obtain the spectra shown in Fig.~\ref{Fig3}(e), which is plotted in the same way as Fig.~\ref{Fig3}(a), where the extracted positions of the upper/lower polaritons, i.e., the ones obtained in Fig.~\ref{Fig3}(a), are shown again for a guide to the eye.  In the bottom-right inset, the polarization of the observed signal when $\lambda_{\mathrm{QD}} - \lambda_{\mathrm{cav}} = 0.5$~nm is shown. The $x$-polarized light is expected to be radiated from the grating outcoupler, because the TE-polarization of the light from the cavity coupled to the waveguide should be maintained. This is verified by the $x$-polarized nature of the signal.  The avoided crossing is visible in the plot, ensuring the realization of the strongly coupled single QD-cavity system accessible via the silicon waveguide on the CMOS platform.  The signal of the QD is still observable when it is slightly off-resonant from the cavity; however, it is no longer visible as the QD-cavity detuning becomes larger, as the QD signal is expected to be visible only through the coupling to the cavity. 

Before summarizing the study, we will briefly discuss the possible improvement of the coupling strength between the QD and cavity and the waveguide-coupling efficiency. First, the QD-cavity coupling strength depends significantly on the position of the QD within the cavity~\cite{Kuruma2016}. The coupling strength is expected to reach at least 93~$\mu$eV according to Ref.~\cite{Ohta2011}.  
To integrate the QD-cavity system with larger coupling strength on silicon photonic chip, the low yield, which is about several percents in this work, should be improved. The device yield is limited mainly by the uncertainty of the wavelength and position of the QD and the wavelength and quality factor of the cavity. The transfer-printing method permits back-end assembly after individual optimization and pre-screening of the desired elements to address the above issues. This leads to deterministic integration of QD-cavity systems with larger coupling strength and quality factor of the cavity.

The waveguide-coupling efficiency can be estimated from the comparison of the loaded quality factor of the cavity with the unloaded one. However, in the current case, the waveguide-coupling rate $\kappa_{\mathrm{WG}} \sim 2\pi \times 5$~GHz determined by the FDTD calculation is small compared to the experimentally quantified total loss rate of the cavity, $\kappa_{\mathrm{tot}} \sim 2\pi \times 32$~GHz, whose sample-to-sample fluctuation hinders the relevant estimation of $\kappa_{\mathrm{WG}}$ in the experiment.  By adopting the above-mentioned values of $\kappa_{\mathrm{WG}}$ and $\kappa_{\mathrm{tot}}$, we get the coupling efficiency $\eta_{\mathrm{exp}} = \kappa_{\mathrm{WG}}/\kappa_{\mathrm{tot}} \sim 16$\,\%.  The small value of $\eta_{\mathrm{exp}}$ not only hinders the relevant estimation of the coupling efficiency as mentioned above, but also limits the signal-to-noise ratio of the spectra in Fig.~\ref{Fig3}(e).  If the unloaded quality factor of the cavity is improved to $8 \times 10^4$~\cite{Kuruma2018} and the waveguide-coupling rate is increased to yield a loaded quality factor of $1 \times 10^4$, when we set the waveguide-resonator gap to be $450$~nm, the coupling efficiency becomes 88\,\% while maintaining the system in the strong coupling regime. These improvements pave the way to efficient optical control and dispersive readout of the quantum states of a QD, and realization of single-photon nonlinear elements even enabling photon-number resolution, where QPIC acquires further functionality and scalability on a CMOS-processed silicon photonic chip.

In conclusion, we demonstrated for the first time a strongly coupled QD-cavity system integrated on a CMOS-processed silicon waveguide. The device was fabricated by the transfer-printing method, which overcame the difficulty of the hybrid integration method. Low-temperature micro-PL spectroscopy verified the strong coupling between the QD and cavity with observation of the vacuum Rabi splitting as well as the waveguide coupling of the cavity. This work serves as an important stepping stone toward the realization of novel quantum devices incorporating single-photon nonlinearity, to be developed on the CMOS platform.

We are grateful to K. Kuruma for fruitful discussion. This work was supported by JSPS KAKENHI Grant-in-Aid for Specially Promoted Research (15H05700), KAKENHI 16K06294 and the New Energy and Industrial Technology Development Organization (NEDO).




\begin{thebibliography}{99}

\bibitem{OBrien2009}
J.~L.~O'Brien, A.~Furusawa, and J.~Vuckovic Nat. Photonics \textbf{3}, 687 (2009).
\bibitem{Silverstone2016}
J.~W.~Silverstone, D.~Bonneau, J.~L.~O'Brien, and M.~G.~Thompson, IEEE J. Sel. Top. Quantum Electron. \textbf{22}, 6700113 (2016).
\bibitem{NielsenChuang}
M.~A.~Nielsen and I.~L.~Chuang, \textit{Quantum Computation and Quantum Information} (Cambridge University Press, Cambridge, 2010).
\bibitem{Georgescu2014}
I.~M.~Georgescu, S.~Ashhab, and F.~Nori, Rev. Mod. Phys. \textbf{86}, 153 (2014).
\bibitem{Gisin2002}
N.~Gisin, G.~Ribordy, W.~Tittel, and H.~Zbinden, Rev. Mod. Phys. \textbf{74}, 145 (2002).
\bibitem{Degen2017}
C.~ L.~Degen, F.~Reinhard, and P.~Cappellaro, Rev. Mod. Phys. \textbf{89}, 035002 (2017).
\bibitem{Soref2006}
R.~Soref, IEEE J. Sel. Top. Quantum Electron. \textbf{12}, 1678 (2006).
\bibitem{Dai2012}
D.~Dai, J.~Bauters and J.~E.~Bowers, Light: Science and Applications \textbf{1}, e1 (2012).
\bibitem{Raimond2001}
J.~M.~Raimond, M.~Brune, and S.~Haroche, Rev.~Mod.~Phys.~\textbf{73}, 565 (2001).
\bibitem{Aoki2006}
T.~Aoki, B.~Dayan, E.~Wilcut, W.~P.~Bowen, A.~S.~Parkins, T.~J.~Kippenberg, K.~J.~Vahala, and H.~J.~Kimble, Nature \textbf{443}, 671 (2006).
\bibitem{Steiner2013}
M.~Steiner, H.~M.~Meyer, C.~Deutsch, J.~Reichel, and M.~K\"{o}hl, Phys. Rev. Lett. \textbf{110}, 043003 (2013).
\bibitem{Park2006}
Y.-S.~Park, A.~K.~Cook, and H.~Wang, Nano Lett. \textbf{6} 2075 (2006).
\bibitem{Xiang2013}
Z.-L.~Xiang, S.~Ashhab, J.~Q.~You, and F.~Nori, Rev. Mod. Phys. \textbf{85}, 623 (2013).
\bibitem{Yoshie2004}
T.~Yoshie, A.~Scherer, J.~Hendrickson, G.~Khitrova, H.~Gibbs, G.~Rupper, C.~Ell, O.~Shchekin, and D.~Deppe, Nature (London) \textbf{432}, 200 (2004).
\bibitem{Reithmaier2004}
J.~P.~Reithmaier, G.~Sek, A.~L\"{o}ffler, C.~Hofmann, S.~Kuhn, S.~Reitzenstein, L.~V.~Keldysh, V.~D.~Kulakovskii, T.~L.~Reinecke, and A.~Forchel, Nature (London) \textbf{432}, 197 (2004).
%
 
\bibitem{Michler2000}
P.~Michler, A.~Kiraz, C.~Becher, W.~V.~Schoenfeld, P.~M.~Petreff, L.~Zhang, E.~Hu, and A.~Imamoglu, Science \textbf{290}, 2282 (2000).
\bibitem{Kuhn2002}
A.~Kuhn, M.~Hennrich, and G.~Rempe, Phys. Rev. Lett, \textbf{89}, 067901 (2002).
\bibitem{Englund2010}
D.~Englund, B.~Shields, K.~Rivoire, F.~Hatami, J.~Vuckovic, H.~Park, and M.~D.~Lukin, Nano Lett. \textbf{10}, 3922 (2010).
\bibitem{Houck2007}
A.~A.~Houck, D.~I.~Schuster, J.~M.~Gambetta, J.~A.~Schreier, B.~R.~Johnson, J.~M.~Chow, L.~Frunzio, J.~Majer, M.~H.~Devoret, S.~M.~Girvin, and R.~J.~Shoelkopf, Nature \textbf{449}, 328 (2007).
\bibitem{Shomroni2014}
I.~Shomroni, S.~Rosenblum, Y.~Lovsky, O.~Bechler, G.~Guendelman, and B.~Dayan, Science \textbf{345}, 903 (2014).
\bibitem{Volz2012}
T.~Volz, A.~Reinhard, M.~Winger, A.~Badolato, K.~J.~Hennessy, E.~L.~Hu, and A.~Imamoglu, Nat. Photonics \textbf{6}, 605 (2012).
\bibitem{Hoi2011}
T.-C.~Hoi, C.~M.~Wilson, G.~Johansson, T.~Palomaki, B.~Peropadre, and P.~Delsing, Phys. Rev. Lett. \textbf{107}, 073601 (2011).
\bibitem{Tiecke2014}
T.~G.~Tiecke, J.~D.~Thompson, N.~P.~de Leon, L.~R.~Liu, V.~Vuletic, and M.~D.~Lukin, Nature \textbf{508}, 241 (2014).
\bibitem{Javadi2015}
A.~Javadi, I.~S\"{o}llner, M.~Arcari, S.~Lindskov Hansen, L.~Midolo, S.~Mahmoodian, G.~Kirsanske, T.~Pregnolato, E.~H.~Lee, J.~D.~Song, S.~Stobbe, and P.~Lodahl, Nat. Commun. \textbf{6}, 8655 (2015).
\bibitem{Zadeh2016}
I.~E.~Zadeh, A.~W.~Elshaari, K.~D.~J\"{o}ns, A.~Fognini, D.~Dalacu, P.~J.~Poole,
M.~E.~Reimer, and V.~Zwiller, Nano Lett. \textbf{16}, 2289 (2016).
\bibitem{Elshaari2017}
A.~W.~Elshaari, I.~E.~Zadeh, A.~Fognini, M.~E.~Reimer, D.~Dalacu,
P.~J.~Poole, V.~Zwiller, and K.~D.~J\"{o}ns, Nat. Commun. \textbf{8}, 379
(2017).
\bibitem{Kim2017}
J.-H.~Kim, S.~Aghaeimeibodi, C.~J.~K.~Richardson, R.~P.~Leavitt, D.~Englund, and E.~Waks, Nano Lett. \textbf{17}, 7394 (2017).
\bibitem{Davanco2017}
M.~Davanco, J.~Liu, L.~Sapienza, C.-Z.~Zhang, J.~V.~De Miranda Cardoso,
V.~Verma, R.~Mirin, S.~W.~Nam, L.~Liu, and K.~Srinivasan, Nat. Commun.
\textbf{8}, 889 (2017).
\bibitem{Luxmoore2013}
I.~J.~Luxmoore, R.~Toro, O.~Del Pozo-Zamudio, N.~A.~Wasley, E.~A.~Chekhovich, A.~M.~Sanchez, R.~Beanland, A.~M.~Fox, M.~S.~Skolnick, H.~Y.~Liu, and A.~I.~Tartakovskii, Sci. Rep. \textbf{3}, 1239 (2013). 



\bibitem{Menard2004}
E.~Menard, K.~J.~Lee, D.-Y.~Khang, R.~G.~Nuzzo, and J.~A.~Rogers, Appl. Phys. Lett. \textbf{84}, 5398 (2004).
\bibitem{Meitl2006}
M.~A.~Meitl, Z.-T.~Zhu, V.~Kumar, K.~J.~Lee, X.~Feng, Y.~Y.~Huang, I.~Adesida, R.~G.~Nuzzo, and J.~A.~Rogers, Nat. Materials \textbf{5}, 23 (2006).
\bibitem{Yoon2015}
J.~Yoon, S.-M.~Lee, D.~Kang, M.~A.~Meitl, C.~A.~Bower, and J.~A.~Rogers, Adv. Optical Mater. \textbf{3}, 1313 (2015). 

\bibitem{Brune1990}
M.~Brune, S.~Haroche, V.~Lefevre, J.~M.~Raimond, and N.~Zagury, Phys. Rev. Lett. \textbf{65}, 976 (1990).
\bibitem{Wallraff2004}
A.~Wallraff, D.~I.~Schuster, A.~Blais, L.~Frunzio, R.-S.~Huang, J.~Majer, S.~M.~Girvin, and R.~J.~Schoelkopf, Nature \textbf{431}, 162 (2004).

\bibitem{Deotare2009}
P.~B.~Deotare, M.~W.~McCutcheon, I.~W.~Frank, M.~Khan, and M.~Loncar, Appl. Phys. Lett. \textbf{94}, 121106 (2009).
\bibitem{Kuramochi2010}
E.~Kuramochi, H.~Taniyama, T.~Tanabe, K.~Kawasaki, Y.-G.~Roh, and M.~Notomi, Opt. Express \textbf{18}, 15859 (2010).
\bibitem{Gong2010}
Y.~Gong, B.~Ellis, G.~Shambat, T.~Sarmiento, J.~S.~Harris, and J.~Vuckovic, Opt. Express \textbf{18}, 8781 (2010).
\bibitem{Ohta2011}
R.~Ohta, Y.~Ota, M.~Nomura, N.~Kumagai, S.~Ishida, S.~Iwamoto, and Y.~Arakawa, Appl. Phys. Lett. \textbf{98}, 173104 (2011).

\bibitem{Katsumi2018}
R.~Katsumi, Y.~Ota, M.~Kakuda, S.~Iwamoto, and Y.~Arakawa, Optica \textbf{5}, 691 (2018).
\bibitem{AO2018}
A.~Osada, Y.~Ota, R.~Katsumi, K.~Watanabe, S.~Iwamoto, and Y.~Arakawa, Appl. Phys. Express \textbf{11}, 072002 (2018). 


\bibitem{Hennessy2007}
K.~Hennessy, A.~Badolato, M.~Winger, D.~Gerace, M.~Atat\"{u}re, S.~Gulde, S.~F\"{a}lt, E.~L.~Hu, and A.~Imamoglu, Nature (London) \textbf{445}, 896 (2007).
\bibitem{JC1963}
E.~T.~Jaynes and F.~W.~Cummings, Proceedings of the IEEE \textbf{51}, 89 (1963).
\bibitem{Kuruma2016}
K.~Kuruma, Y.~Ota, M.~Kakuda, D.~Takamiya, S.~Iwamoto, and Y.~Arakawa, Appl. Phys. Lett. \textbf{109}, 071110 (2016).
\bibitem{Kuruma2018}
K.~Kuruma, Y.~Ota, M.~Kakuda, S.~Iwamoto, and Y.~Arakawa, Phys. Rev. B \textbf{97}, 235448 (2018).


\end{thebibliography}
\end{document}